\documentclass[lettersize,journal]{IEEEtran}
\usepackage{amsmath,amsfonts}
\usepackage{algorithmic}
\usepackage{algorithm}
\usepackage{array}
\usepackage[caption=false,font=normalsize,labelfont=sf,textfont=sf]{subfig}
\usepackage{textcomp}
\usepackage{stfloats}
\usepackage{url}
\usepackage{verbatim}
\usepackage{graphicx}
\usepackage{cite}
\usepackage{booktabs}
\usepackage{multirow}
\usepackage{rotating}
\usepackage{pifont}
\usepackage{booktabs}
\usepackage{tabularx}
\usepackage{multirow}
\usepackage{xcolor}
\usepackage{makecell}
\usepackage{placeins}

\usepackage[hidelinks]{hyperref}
\renewcommand{\baselinestretch}{0.980}

\begin{document}
\title{MusicMark: A Robust Generative Watermarking Framework for Music Generation}

\author{
Seohwan Yun,
Jeeyoung Yun,
Yongjin Kim,
Juyeon Lee,
and Sungwoong Kim
\thanks{
This work has been submitted to the IEEE for possible publication.
Copyright may be transferred without notice, after which this version may no longer be accessible.
}
\thanks{
Seohwan Yun, Jeeyoung Yun, Yongjin Kim and Sungwoong Kim are with the Department of Artificial Intelligence, Korea University, Seoul, Republic of Korea.
}
\thanks{
Juyeon Lee is with the Department of Computer Engineering, Inha University, Incheon, Republic of Korea.
}%
\thanks{
Corresponding author: Sungwoong Kim
(e-mail: swkim01@korea.ac.kr).
}}

\markboth{PREPRINT}%
{Yun \MakeLowercase{\textit{et al.}}: MusicMark}


\maketitle

\begin{abstract}
AI music generation has rapidly advanced alongside commercial platforms, raising the need for reliable watermarking for provenance and attribution. However, existing audio watermarking research has largely focused on speech, and applying speech-oriented methods to music is challenging due to music's complex structure and rich acoustic texture. Most existing methods are post-hoc, adding imperceptible perturbations after generation rather than embedding watermarks as part of the content. This makes them fragile under various transformations and especially vulnerable to neural codec re-synthesis, which can discard imperceptible residual signals. Moreover, since generation and watermarking are decoupled, the watermarking step can be bypassed or omitted, weakening provenance guarantees. To address these issues, we propose MusicMark, which, to the best of our knowledge, is the first generative watermarking framework for music. Specifically, MusicMark embeds watermark messages into the semantic latent space during generation, incorporating the watermark as part of the musical content and ensuring robustness against diverse attacks, particularly neural codec re-synthesis. To this end, we introduce a watermark adapter into a diffusion-based generation model to embed watermark messages across denoising steps. The adapter and detector are trained with a joint objective that preserves fidelity by constraining watermarked latents close to their unwatermarked reference latents, while improving robustness through comprehensive attack augmentations. Experiments demonstrate that MusicMark substantially outperforms post-hoc baselines across diverse attacks including neural codec re-synthesis, while maintaining comparable generation quality. We further introduce a cover-song attack, converting the singing voice while preserving musical content, and show that MusicMark remains more robust than post-hoc methods.
\end{abstract}


\section{Introduction}
\IEEEPARstart{R}{ecent} advances in open-source music generation models~\cite{lei2026levo, gong2025ace, ning2025diffrhythm, jiang2025diffrhythm, yuan2025yue, yang2026heartmula, liu2025jam, evans2026stable} and commercial AI music platforms~\cite{SunoAI,UdioAI} have expanded AI content generation beyond speech synthesis, allowing users to create and distribute high-quality music from text prompts, lyrics, and other high-level conditions. As such content rapidly proliferates, reliable music watermarking becomes increasingly important for verifying the provenance and attribution of generated music, including its origin and source model.

Existing audio watermarking research has largely focused on speech~\cite{sanroman2024proactive, singh2024silentcipher, chen2023wavmark, timbrewatermarking-ndss2024, liu2023dear, zhou2024traceablespeechproactivelytraceabletexttospeech, cheng2024hifi, liu2026vocbulwark}. Applying these speech-oriented methods to music is challenging, as music differs substantially from speech in signal characteristics, musical structure, and perceptual constraints. 
Music typically spans a broader frequency range, uses higher sampling rates, and contains polyphonic mixtures of instruments and vocals organized through melody, harmony, rhythm, and long-range temporal dependencies~\cite{li2024jen, agostinelli2023musiclm, copet2023simple}. 
Moreover, since human listeners are highly sensitive to musical dissonance and small melodic or harmonic errors, even subtle watermark-induced perturbations can degrade perceived musical quality~\cite{li2024jen, copet2023simple}. 
Music representations also rely heavily on pitch, tonality, harmony, and timbre~\cite{li2024mert}. 
Consequently, watermarking music requires preserving rich perceptual details while maintaining robustness to diverse audio transformations.

Most existing audio watermarking methods are post-hoc~\cite{sanroman2024proactive, singh2024silentcipher, chen2023wavmark, timbrewatermarking-ndss2024, liu2023dear}, inserting imperceptible perturbations into already generated audio. In this setting, the watermark is inserted to preserve perceptual quality, rather than being semantically or structurally integrated with the underlying audio content. As a result, these watermark signals can be fragile to temporal changes, frequency-domain transformations, and audio compression, which may remove, misalign, or distort the inserted signal. This limitation becomes especially pronounced under neural codec re-synthesis~\cite{defossez2022high, siuzdak2024snac, kumar2023high}, where codecs preserve semantic content while discarding signals not captured in the semantic representation. This vulnerability has been widely identified as a fundamental limitation of post-hoc audio watermarking~\cite{OezerChoiEtAl25_RAWBench_Interspeech, wen2025sok, o2025deep, liu2024audiomarkbench}. Furthermore, because generation and watermarking are performed separately, the post-hoc watermarking step can be bypassed or ignored, weakening provenance guarantees~\cite{san2025latent, lee2025robust}.

Generative watermarking mitigates these issues by embedding watermarks during audio generation rather than as a separate post-hoc step~\cite{cheng2024hifi, liu2024groot, liu2026vocbulwark}. Existing generative audio watermarking methods generally embed the watermark at the decoder or vocoder stage during waveform rendering, after the semantic latent representation or intermediate acoustic content has already been formed. This late-stage insertion leaves the watermark weakly aligned with the generated content and vulnerable to downstream transformations, especially neural codec re-synthesis.
 
To address these limitations, we propose \textbf{MusicMark}, a generative watermarking framework for music that embeds watermark messages directly into the \textbf{semantic latent space}  before audio rendering.
In particular, we design a watermark adapter for a diffusion-based music generation model, where a separate cross-attention module conditions the latent denoising process on the watermark message, allowing the message to be embedded as the semantic latent representation is generated.
The resulting watermarked music is then passed to a watermark detector, which identifies watermark presence and extracts the embedded message.

To train the watermark adapter and detector built on top of a base music generation model, we introduce a joint objective with latent generation, fidelity, and watermark losses. In particular, we add a latent consistency loss to constrain the distance between watermarked latents and their unwatermarked references, preserving generation quality while the watermark loss enables robust message embedding. We further incorporate diverse attack augmentations during training to enhance robustness.

Extensive experiments show that MusicMark significantly improves the robustness of watermark detection and message extraction over post-hoc watermarking baselines against diverse attacks, including neural codec re-synthesis, without compromising generation quality. We further introduce a \textbf{cover-song attack}, a music-specific attack that converts the original singing voice to a different singer identity while preserving the song's melody, harmony, lyrics, and instrumental accompaniment. MusicMark maintains reliable detection and message extraction while preserving perceptual quality under this challenging setting, demonstrating that coupling watermarks with the musical semantic representation yields stronger robustness than post-hoc or decoder-level approaches.

Our contributions can be summarized as follows:
\begin{itemize}
    \item We develop MusicMark, the first generative watermarking framework for lyrics- and text-conditioned music generation.
    \item A watermark adapter with a separate cross-attention module is designed and inserted into a diffusion-based latent denoising model, enabling watermark messages to be effectively embedded in the semantic space of generated music.
    \item A joint training strategy is developed for the watermark adapter and detector, combining latent generation, fidelity, and watermark losses with attack augmentations, including a latent consistency loss that preserves generation quality by aligning watermarked latents with unwatermarked references.
    \item MusicMark is empirically shown to significantly outperform post-hoc watermarking baselines under diverse audio transformations, especially neural codec re-synthesis and cover-song transformations, while preserving perceptual quality.
\end{itemize}

\begin{figure*}[t!]
    \centering
    \includegraphics[width=\textwidth]{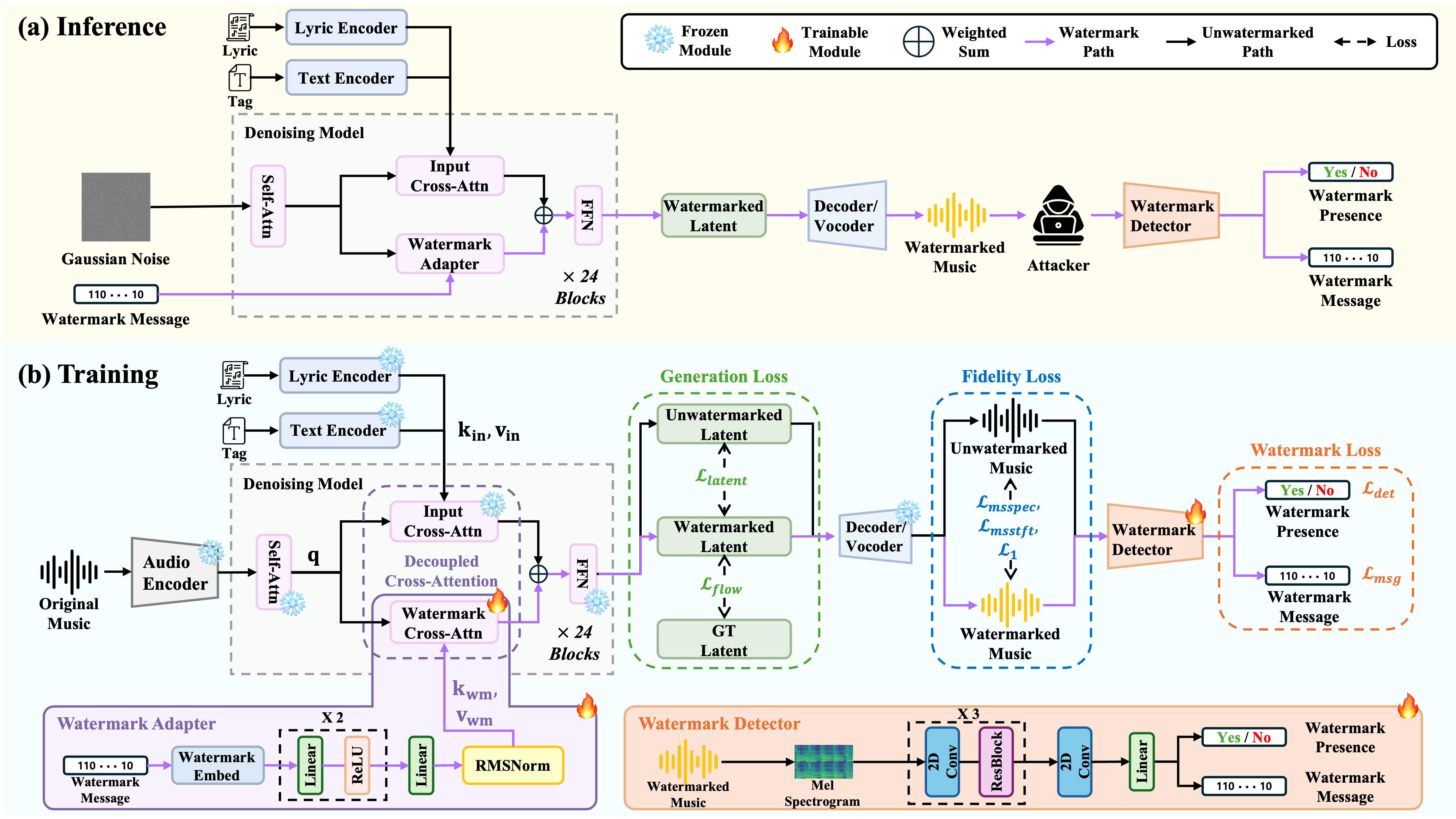}
    \caption{Overview of MusicMark. a) Inference process, where lyrics and style tags are used as text conditions, and the watermark adapter injects the watermark message via decoupled cross-attention to generate the watermarked music latent. The decoder/vocoder renders this latent into watermarked music. The detector then detects watermark presence and extracts the embedded message from the watermarked music, even after diverse attacks. b) Training process, where the base generation model is frozen and only the watermark adapter and detector are trained with generation, fidelity, and watermark losses for robust watermark detection and quality preservation.}
    \label{overview}
\end{figure*}

\section{Related Works}

\subsection{Post-hoc Audio Watermarking}
Post-hoc audio watermarking embeds messages into already generated audio as imperceptible perturbations. WavMark~\cite{chen2023wavmark} uses an invertible neural network for message embedding and extraction, AudioSeal~\cite{sanroman2024proactive} supports localized watermark detection, and SilentCipher~\cite{singh2024silentcipher} incorporates psychoacoustic masking for imperceptible audio watermarking. Other methods focus on speech-specific settings, such as timbre-based watermarking for voice-cloning detection~\cite{timbrewatermarking-ndss2024} and watermarking methods robust to re-recording attacks~\cite{liu2023dear}.
Although post-hoc methods are flexible and can be applied to existing audio, they insert watermarks only after the audio content has already been formed. Thus, the watermark remains an external signal rather than part of the content-forming representation and can be weakened by diverse attacks~\cite{wen2025sok}. This issue is especially pronounced under neural codec re-synthesis, a widely observed failure case for post-hoc audio watermarking~\cite{OezerChoiEtAl25_RAWBench_Interspeech,o2025deep,liu2024audiomarkbench}. Moreover, because post-hoc watermarking requires a separate watermarking step after generation, the watermark can be bypassed if this post-processing stage is omitted.

\subsection{Generative Audio Watermarking}
Generative audio watermarking aims to embed watermarks during generation rather than as a post-hoc perturbation. Model-level approaches modify the generator itself, either by embedding watermarks into model weights~\cite{feng2025robust}, fine-tuning speech generation models or vocoders~\cite{zhou2024traceablespeechproactivelytraceabletexttospeech, cheng2024hifi}, or training audio generators on pre-watermarked data~\cite{san2025latent}. Input-level approaches inject an encoded watermark latent into the model input~\cite{liu2024groot, lee2025robust}. Additional-parameter injection methods keep the base generator frozen and inject watermarks through lightweight trainable modules into intermediate acoustic representations~\cite{liu2026vocbulwark}. However, model-level approaches require fine-tuning or retraining the generator, which can be costly and may affect generation quality.
Input-level and additional-parameter injection methods may inject watermarks into latent or intermediate representations, but they typically operate on already formed representations rather than during the process in which semantic latents are generated. As a result, the watermark can remain weakly coupled with the semantic content and vulnerable to neural codec re-synthesis and voice conversion.
\section{Methods}
\subsection{Preliminaries}
\label{sec:prelim}
Recent progress in high-quality music generation has been driven by diffusion-based models~\cite{gong2025ace,ning2025diffrhythm,jiang2025diffrhythm,evans2026stable}. Unlike autoregressive models~\cite{lei2026levo, yuan2025yue, liu2025songgen} that generate audio tokens sequentially, diffusion-based models denoise audio tokens in parallel, enabling fast and high-quality music generation. Diffusion-based models also provide greater controllability over the
generated output than autoregressive models, enabling more effective
embedding of watermark messages into the generation process. These models typically generate latent representations of music rather than waveforms directly: an encoder maps music into a latent space, and a decoder or vocoder reconstructs the waveform from the resulting latent representation. The generation process starts from Gaussian noise and iteratively denoises it into a clean latent via the denoising model $\mathbf{v}_\theta$. Text conditions such as lyrics and style tags guide this process through cross-attention. 

\subsection{Overview}
We propose \textbf{MusicMark}, a generative music watermarking framework that embeds watermark information into the semantic latent space during music generation. Specifically, the denoising model $\mathbf{v}_\theta$ takes Gaussian
noise as input and iteratively denoises it into a watermarked latent, guided by text prompts and a watermark message through two parallel cross-attention modules. The input cross-attention module incorporates text prompts such as lyrics and style tags, while the watermark cross-attention module injects the watermark message embedded through the adapter. The outputs of the two modules are combined via a weighted sum, seamlessly embedding the watermark into the semantic latent throughout the denoising process. The decoder/vocoder then converts the watermarked latent into watermarked music. Given the resulting watermarked music, the watermark detector identifies watermark presence and extracts the embedded message, even under diverse attacks. This inference process is illustrated in Fig.~\ref{overview}(a). During training, the base generation model is frozen, and only the watermark adapter and detector are optimized, as illustrated in Fig.~\ref{overview}(b).

In the following subsections, we describe the watermark adapter (Section~\ref{sec:watermark_adapter}), music decoder/vocoder (Section~\ref{sec:music_decoder_vocoder}), watermark detector (Section~\ref{sec:watermark_detector}), and training strategy, which includes attack augmentation and training objectives (Section~\ref{sec:train_strategy}).

\subsection{Watermark Adapter}
\label{sec:watermark_adapter}
The watermark adapter $A$ injects a watermark message into the semantic latents during music generation. It consists of two components: a watermark embedding module that maps the multi-bit binary message to a watermark representation, and a decoupled cross-attention module that incorporates this representation into the semantic latents.

\noindent\textbf{Watermark Embedding.}
Given an $N$-bit binary watermark message $\mathbf{m}\in\{0,1\}^{N}$, we assign two learnable embeddings to each bit position, corresponding to bit values 0 and 1. This yields an embedding table with $2N$ entries. For the $i$-th bit, the embedding index is defined as
\begin{equation}
    I_i = 2i + m_i, \quad i=0,\ldots,N-1.
\end{equation}
Using the index vector $\mathbf{I}=[I_0,\ldots,I_{N-1}]$, we select one embedding for each bit position and obtain a sequence of position-specific message embeddings. The selected embeddings are then projected and normalized to form the watermark representation:
\begin{equation}
    \mathbf{Z} = \mathrm{RMSNorm}(f_{\mathrm{Proj}}(\mathrm{Emb}(\mathbf{I}))) \in \mathbb{R}^{N\times d},
\end{equation}
where $\mathrm{Emb}(\cdot)$ denotes a lookup into the embedding table, $f_{\mathrm{Proj}}(\cdot)$ is an MLP projector consisting of two Linear--ReLU blocks followed by a final linear layer, and $d$ is the latent feature dimension of the denoising model. 

\noindent\textbf{Decoupled Cross-Attention.}
We inject the watermark representation $\mathbf{z}$ through a separate watermark cross-attention module added alongside the original text-conditioning cross-attention module. By keeping these two modules separate, the original generation quality of the base model is preserved while allowing the watermark message to be injected without interfering with the existing text conditioning. Let $\mathbf{x}\in\mathbb{R}^{S\times d}$ denote the semantic latent features at a cross-attention layer, where $S$ is the latent sequence length. Let $\mathbf{c}\in\mathbb{R}^{L\times d}$ denote the text condition embeddings encoded from the lyrics and style tags, where $L$ is the condition sequence length.

The original input cross-attention module computes cross-attention from the latent features to the text condition embeddings:
\begin{equation}
    \mathbf{q} = \mathbf{x}\mathbf{W}_{Q}, \quad
    \mathbf{k}_{\mathrm{in}} = \mathbf{c}\mathbf{W}_{K}, \quad
    \mathbf{v}_{\mathrm{in}} = \mathbf{c}\mathbf{W}_{V},
\end{equation}
\begin{equation}
    \mathbf{h}_{\mathrm{in}} =
    \mathrm{Attn}(\mathbf{q}, \mathbf{k}_{\mathrm{in}}, \mathbf{v}_{\mathrm{in}}),
\end{equation}
where $\mathbf{W}_{Q}$, $\mathbf{W}_{K}$, and $\mathbf{W}_{V}$ are the frozen projection weights of the base denoising model.
The watermark cross-attention module reuses the same latent query $\mathbf{q}$, but derives its key and value from the watermark representation $\mathbf{z}$:
\begin{equation}
    \mathbf{k}_{\mathrm{wm}} = \mathbf{z}\mathbf{W}^{\mathrm{wm}}_{K}, \quad
    \mathbf{v}_{\mathrm{wm}} = \mathbf{z}\mathbf{W}^{\mathrm{wm}}_{V},
\end{equation}
\begin{equation}
    \mathbf{h}_{\mathrm{wm}} =
    \mathrm{Attn}(\mathbf{q}, \mathbf{k}_{\mathrm{wm}}, \mathbf{v}_{\mathrm{wm}}).
\end{equation}
The outputs of the two modules are then combined via a weighted sum:
\begin{equation}
    \mathbf{h}_{\mathrm{out}} =
    \mathbf{h}_{\mathrm{in}} + \alpha \mathbf{h}_{\mathrm{wm}},
\end{equation}
where $\alpha$ is a learnable scale. We initialize $\mathbf{W}^{\mathrm{wm}}_{K}$ and $\mathbf{W}^{\mathrm{wm}}_{V}$ to zero so that the watermark branch is inactive at the beginning of training, i.e., $\mathbf{h}_{\mathrm{out}}=\mathbf{h}_{\mathrm{in}}$. As training proceeds, the watermark cross-attention module gradually learns to inject the watermark message into the latent without affecting the original input cross-attention module. This watermark adapter is inserted into selected cross-attention layers of the base generation model, and we analyze the effect of this choice in Section~\ref{sec:ablation}.

\subsection{Music Decoder/Vocoder}
\label{sec:music_decoder_vocoder}
The music decoder/vocoder $\mathcal{V}$ maps a denoised semantic latent to a stereo waveform. For the watermarked latent prediction $\hat{\mathbf{x}}_0^{\mathrm{wm}}\in\mathbb{R}^{S\times d}$, this rendering process is written as:
\begin{equation}
    \hat{\mathbf{w}}^{\mathrm{wm}}
    =
    \mathcal{V}(\hat{\mathbf{x}}_0^{\mathrm{wm}})
    \in \mathbb{R}^{2\times T},
\end{equation}
where $T$ denotes the waveform length. We keep $\mathcal{V}$ frozen throughout training, therefore it serves only as the audio rendering module.

\subsection{Watermark Detector}
\label{sec:watermark_detector}
The watermark detector $\mathcal{D}$ identifies watermark presence and extracts the embedded message. It converts $\hat{\mathbf{w}}^{\mathrm{wm}}$ into a log-mel spectrogram and feeds it to a detector feature extractor $\phi_{\mathrm{det}}$, initialized from the pretrained audio encoder. This produces a latent-aligned feature map:
\begin{equation}
    \mathbf{F}
    =
    \phi_{\mathrm{det}}\!\left(\mathrm{Mel}(\hat{\mathbf{w}}^{\mathrm{wm}})\right)
    \in \mathbb{R}^{S \times C_{\mathrm{det}}\times F_{\mathrm{det}}},
\end{equation}
where $C_{\mathrm{det}}$ and $F_{\mathrm{det}}$ denote the channel and frequency dimensions, respectively, and $S$ matches the semantic latent sequence length. This temporal alignment allows the detector to predict watermark presence and message bits at each latent-aligned time step.
A prediction head $h_{\mathrm{pred}}$ then collapses the frequency dimension and projects the channels to $2+N$ logits:
\begin{equation}
    \mathbf{R}
    =
    h_{\mathrm{pred}}(\mathbf{F})
    \in \mathbb{R}^{S \times (2+N)}.
\end{equation}
We split $\mathbf{R}=\mathcal{D}(\hat{\mathbf{w}}^{\mathrm{wm}})$ into presence logits $\mathbf{R}_p\in\mathbb{R}^{S\times 2}$ and message logits $\mathbf{R}_m\in\mathbb{R}^{S\times N}$.

\subsection{Training}
\label{sec:train_strategy}
We train the watermark adapter and detector while keeping the base music generation model frozen. Our training objective balances three goals: keeping the watermarked latent close to its unwatermarked reference, preserving perceptual music quality, and enabling robust watermark detection and message extraction. Accordingly, we exploit a latent generation loss, a fidelity loss, and a watermark loss.

\noindent\textbf{Attack Augmentations.}
To improve the robustness of watermark detection, we apply attack augmentations to the watermarked music during training before passing it to the detector. We use 10 training attacks: Gaussian noise, low/high-pass filtering, echo, resampling, speed change, MP3 compression, EnCodec re-synthesis, and two music-specific cover-song attacks, namely vocal-conversion cover song and cover song with cut. Since the watermark adapter and detector are trained jointly, we handle non-differentiable attacks such as MP3 compression, neural codec re-synthesis, and cover-song conversion with the straight-through estimator~\cite{yin2019understanding}: the attacked music is used in the forward pass, while gradients are passed to the original watermarked music in the backward pass. Training details and attack configurations are provided in Appendices~\ref{app:imple_detail} and~\ref{app:attack_details}, respectively.

\noindent\textbf{Latent Generation Loss.}
The latent generation loss ensures that music generation quality is maintained during watermark injection. It consists of a flow matching loss and a latent consistency loss. Among several denoising objectives used in diffusion-based generation, we adopt flow matching, which has recently been widely used for efficient high-quality generation. We further introduce the latent consistency loss as, to the best of our knowledge, the first regularization term designed to align watermarked and unwatermarked latents for generative music watermarking. The flow matching loss encourages accurate velocity prediction even as the watermark message is embedded into the latent, while the latent consistency loss keeps the watermarked latent close to the unwatermarked latent obtained using the same text conditions without watermark conditioning. Specifically, the flow matching loss is defined as:
\begin{equation}
\mathcal{L}_{\text{flow}} = \mathbb{E}_{\mathbf{x}_0, \boldsymbol{\epsilon},
t}\left[\left\|\mathbf{v}_\theta(\mathbf{x}_t, t, \mathbf{c}, \mathbf{m}) -
(\boldsymbol{\epsilon} - \mathbf{x}_0)\right\|_2^2\right],
\end{equation}
where $\mathbf{x}_0$ is the clean latent, $\boldsymbol{\epsilon} \sim
\mathcal{N}(\mathbf{0}, \mathbf{I})$ is Gaussian noise, $t \sim
\mathcal{U}[0,1]$ is the time step, $\sigma_t \in [0,1]$ is the noise
schedule, and $\mathbf{x}_t = (1-\sigma_t)\mathbf{x}_0 +
\sigma_t\boldsymbol{\epsilon}$ is the noisy latent at time $t$. For the latent consistency loss, the denoised latents are obtained via:
\begin{align}
\hat{\mathbf{x}}_0^{wm} &= -\sigma_t \mathbf{v}_\theta(\mathbf{x}_t, t,
\mathbf{c}, \mathbf{m}) + \mathbf{x}_t, \\
\hat{\mathbf{x}}_0 &= -\sigma_t \mathbf{v}_\theta(\mathbf{x}_t, t,
\mathbf{c}) + \mathbf{x}_t,
\end{align}
where $\hat{\mathbf{x}}_0^{wm}$ and $\hat{\mathbf{x}}_0$ are the watermarked
and unwatermarked denoised latents, respectively. The latent consistency
loss treats the unwatermarked latent as a fixed target via a stop-gradient
$\mathrm{sg}(\cdot)$:
\begin{equation}
\mathcal{L}_{\mathrm{latent}} = \mathbb{E}_{\mathbf{x}_0, \boldsymbol{\epsilon},
t}\left[\left\|\hat{\mathbf{x}}_0^{wm} -
\mathrm{sg}(\hat{\mathbf{x}}_0)\right\|_2^2\right].
\end{equation}
If this consistency loss were minimized to zero, it would remove watermark information from the latent. However, in the joint objective, it is balanced by the watermark loss, which encourages the detector to extract the embedded message. Thus, the latent consistency loss acts as a regularizer that limits excessive changes to the unwatermarked latent while allowing the adapter to embed watermark information.
The combined generative loss is a weighted sum:
\begin{equation}
\mathcal{L}_{\text{gen}} = \lambda_{\text{flow}}\mathcal{L}_{\text{flow}} +
\lambda_{\text{latent}}\mathcal{L}_{\text{latent}},
\end{equation}
where $\lambda_{\text{flow}}$ and $\lambda_{\text{latent}}$ are loss weights.

\noindent\textbf{Fidelity Loss.}
To maintain perceptual quality, we also compare the watermarked waveform $\hat{\mathbf{w}}^{\mathrm{wm}} = \mathcal{V}(\hat{\mathbf{x}}_0^{wm})$ with an unwatermarked waveform $\hat{\mathbf{w}} = \mathcal{V}(\hat{\mathbf{x}}_0)$ rendered from the same noise and conditions, where $\mathcal{V}$ is the frozen decoder/vocoder. This isolates the distortion introduced by the watermark embedding. 
The fidelity loss consists of multi-scale mel-spectrogram, multi-scale Short Time Fourier Transform (STFT), and waveform-domain terms:
\begin{equation}
\resizebox{\columnwidth}{!}{$
\mathcal{L}_{\text{msspec}} =
\frac{1}{K}\sum_{k=1}^{K}
\left\|
\log(\mathrm{Mel}_k(\hat{\mathbf{w}}^{\mathrm{wm}})+\delta)
-
\log(\mathrm{Mel}_k(\hat{\mathbf{w}})+\delta)
\right\|_1
$}
\end{equation}
where $\mathrm{Mel}_k(\cdot)$ denotes the mel-spectrogram transform at the $k$-th scale, and $\delta > 0$ is a small constant for numerical stability.

\begin{equation}
\begin{aligned}
\mathcal{L}_{\text{msstft}}
= \sum_{j=1}^{M} \Biggl(
  &\frac{
  \bigl\||G_j(\hat{\mathbf{w}}^{\mathrm{wm}})| - |G_j(\hat{\mathbf{w}})|\bigr\|_{F}
  }{
  \bigl\||G_j(\hat{\mathbf{w}})|\bigr\|_{F}
  } \\
  &+
  \bigl\|\log |G_j(\hat{\mathbf{w}}^{\mathrm{wm}})| - \log |G_j(\hat{\mathbf{w}})|\bigr\|_1
\Biggr),
\end{aligned}
\end{equation}
where $G_j(\cdot)$ denotes the STFT at the $j$-th resolution scale and $\|\cdot\|_{F}$ is the Frobenius norm.

The direct waveform-matching loss is also used as
\begin{equation}
\mathcal{L}_{1} = \left\|\hat{\mathbf{w}}^{\mathrm{wm}}-\hat{\mathbf{w}}\right\|_1.
\end{equation}
The combined fidelity loss is then defined as:
\begin{equation}
\mathcal{L}_{\text{fid}}
=
\lambda_{\text{msspec}}\mathcal{L}_{\text{msspec}}
+
\lambda_{\text{msstft}}\mathcal{L}_{\text{msstft}}
+
\lambda_{1}\mathcal{L}_{1},
\end{equation}
where $\lambda_{\text{msspec}}$, $\lambda_{\text{msstft}}$, and $\lambda_{1}$ are loss weights.

\noindent\textbf{Watermark Loss.}
The watermark loss trains the detector $\mathcal{D}$ to identify watermark presence and recover the embedded message bits. Applying
$\mathcal{D}$ to the watermarked waveform $\hat{\mathbf{w}}^{\mathrm{wm}}$ and the
unwatermarked waveform $\hat{\mathbf{w}}$ yields presence logits
$\mathbf{R}_p^{\mathrm{wm}}, \mathbf{R}_p^{\mathrm{cl}} \in \mathbb{R}^{S \times 2}$
and message logits $\mathbf{R}_m^{\mathrm{wm}} \in \mathbb{R}^{S \times N}$, as
defined in Section~\ref{sec:watermark_detector}.

For detection, let $\mathbf{r}_s^{\mathrm{wm}}, \mathbf{r}_s^{\mathrm{cl}} \in
\mathbb{R}^{2}$ denote the presence logits at latent sequence position $s$, i.e.,
the $s$-th columns of $\mathbf{R}_p^{\mathrm{wm}}$ and $\mathbf{R}_p^{\mathrm{cl}}$.
We train the detector with a per-position cross-entropy loss over the two presence
classes, where label $1$ indicates watermark presence and label $0$ indicates absence:
\begin{equation}
\mathcal{L}_{\mathrm{det}} = \frac{1}{2S}\sum_{s=1}^{S}
\left[
\mathrm{CE}\!\left(\mathbf{r}_s^{\mathrm{wm}}, y=1\right)
+
\mathrm{CE}\!\left(\mathbf{r}_s^{\mathrm{cl}}, y=0\right)
\right],
\end{equation}
where $\mathrm{CE}(\mathbf{r},y)$ denotes the cross-entropy between the
two-class logits $\mathbf{r}\in\mathbb{R}^{2}$ and the target class label
$y\in\{0,1\}$.

For message extraction, only the watermarked waveform is used. Let
$\mathbf{R}_{m,i,s}^{\mathrm{wm}}$ denote the detector logit for the $i$-th
bit at position $s$. We apply the logistic sigmoid to obtain the predicted
bit probability $\hat{m}_{i,s}=\sigma(\mathbf{R}_{m,i,s}^{\mathrm{wm}})$. We then minimize the bit-level binary cross-entropy:
\begin{equation}
\mathcal{L}_{\mathrm{msg}} = -\frac{1}{NS}\sum_{i=1}^{N}\sum_{s=1}^{S}
\Bigl(m_i \log \hat{m}_{i,s} + (1 - m_i)\log(1 - \hat{m}_{i,s})\Bigr),
\end{equation} 
where $m_i \in \{0,1\}$ is the $i$-th ground-truth bit. The total watermark
loss is:
\begin{equation}
\mathcal{L}_{\mathrm{wm}} = \lambda_{\mathrm{det}}\mathcal{L}_{\mathrm{det}}
    + \lambda_{\mathrm{msg}}\mathcal{L}_{\mathrm{msg}},
\end{equation}
where $\lambda_{\mathrm{det}}$ and $\lambda_{\mathrm{msg}}$ are loss weights.

\noindent\textbf{Total Loss.}
Finally, we combine the above loss terms to form the overall training objective:
\begin{equation}
\mathcal{L}
=
\mathcal{L}_{\mathrm{gen}}
+
\mathcal{L}_{\mathrm{fid}}
+
\mathcal{L}_{\mathrm{wm}}.
\end{equation}
The base music generation model remains frozen throughout training, and only the watermark adapter $A$ and detector $\mathcal{D}$ are optimized.
\section{Experiments}
\subsection{Experimental Setup}
\label{sec:experimental_setup}

\noindent\textbf{Baselines.}
We build MusicMark on ACE-Step~\cite{gong2025ace}, which follows this latent diffusion paradigm and provides an open-source, high-quality music generation backbone. We compare MusicMark against post-hoc audio watermarking methods operating at 16 kHz, including WavMark~\cite{chen2023wavmark} and AudioSeal~\cite{sanroman2024proactive}, and 44.1 kHz methods, including SilentCipher~\cite{singh2024silentcipher} and AudioSeal-M. Here, AudioSeal-M extends AudioSeal to 44.1 kHz stereo music and is trained using the same data and attack augmentations as MusicMark for a fair comparison. For WavMark and AudioSeal, which operate at 16 kHz mono, we resample the 44.1 kHz stereo music to 16 kHz and average the two channels before evaluation. Details of training for AudioSeal-M are provided in Appendix~\ref{app:imple_detail}.

For measuring generation quality, we restrict the comparison to 44.1 kHz stereo music to ensure that quality differences are not caused by sampling rate or channel format. Using the same lyrics and tags as text conditions, we compare three types of outputs: original ACE-Step outputs without watermarking, ACE-Step outputs watermarked by the post-hoc methods AudioSeal-M and SilentCipher, and MusicMark outputs produced through generative watermarking. This evaluates the quality impact by watermarking under the same music generation backbone and text conditions.

\noindent\textbf{Training and Evaluation Datasets.}
We train on 50K English music--text pairs from Muse~\cite{jiang2026muse}, totaling approximately 625 hours of music. For evaluation, we construct a 2K evaluation set of lyric--prompt pairs from the public Suno metadata source\footnote{\url{https://huggingface.co/datasets/nyuuzyou/suno}}, enabling balanced sampling across the 20 attack types in watermark robustness evaluation. We use a 300-pair subset of this evaluation set for generation-quality evaluation, selected to preserve diversity in song length, lyrics, and prompts. Details of the dataset construction, filtering procedure, and subset selection are provided in Appendix~\ref{app:dataset}.

\begin{table*}[!t]
\caption{Average watermark decoding performance under no-attack and diverse attack settings. SilentCipher does not have a separate detection head; therefore, its Acc values are marked as ``--''.}
\label{tab:attack_wm}
\centering
\tiny
\setlength{\tabcolsep}{2.0pt}
\renewcommand{\arraystretch}{1.08}
\resizebox{\textwidth}{!}{%
\begin{tabular}{@{}llccc ccc ccc ccc ccc@{}}
\toprule
\multirow{2}{*}{\textbf{Category}} &
\multirow{2}{*}{\textbf{Attack}} &
\multicolumn{3}{c}{\textbf{WavMark}} &
\multicolumn{3}{c}{\textbf{AudioSeal}} &
\multicolumn{3}{c}{\textbf{AudioSeal-M}} &
\multicolumn{3}{c}{\textbf{SilentCipher}} &
\multicolumn{3}{c}{\textbf{MusicMark}} \\
\cmidrule(lr){3-5}
\cmidrule(lr){6-8}
\cmidrule(lr){9-11}
\cmidrule(lr){12-14}
\cmidrule(l){15-17}
& &
\textbf{Acc $\uparrow$} & \textbf{Bit $\uparrow$} & \textbf{Abs $\uparrow$} &
\textbf{Acc $\uparrow$} & \textbf{Bit $\uparrow$} & \textbf{Abs $\uparrow$} &
\textbf{Acc $\uparrow$} & \textbf{Bit $\uparrow$} & \textbf{Abs $\uparrow$} &
\textbf{Acc $\uparrow$} & \textbf{Bit $\uparrow$} & \textbf{Abs $\uparrow$} &
\textbf{Acc $\uparrow$} & \textbf{Bit $\uparrow$} & \textbf{Abs $\uparrow$} \\
\midrule

\multirow{2}{*}{\textbf{Mean}}
& No Attack &
\textbf{1.000} & \textbf{1.000} & \textbf{1.000} &
\textbf{1.000} & 0.987 & 0.838 &
\textbf{1.000} & 0.996 & 0.944 &
-- & \textbf{1.000} & \textbf{1.000} &
\textbf{1.000} & 0.999 & 0.993 \\

& Attack &
0.892 & 0.536 & 0.254 &
0.806 & 0.760 & 0.293 &
0.946 & 0.743 & 0.400 &
-- & 0.730 & 0.586 &
\textbf{0.994} & \textbf{0.987} & \textbf{0.869} \\

\midrule
\multirow{8}{*}{\textbf{Common}}
& Gaussian Noise &
0.987 & 0.970 & 0.955 &
\textbf{1.000} & 0.956 & 0.495 &
\textbf{1.000} & 0.993 & 0.901 &
-- & 0.939 & 0.865 &
\textbf{1.000} & \textbf{1.000} & \textbf{1.000} \\

& Reverb &
\textbf{1.000} & 0.993 & 0.929 &
0.920 & 0.796 & 0.089 &
\textbf{1.000} & 0.832 & 0.143 &
-- & \textbf{1.000} & \textbf{1.000} &
0.996 & 0.980 & 0.741 \\

& Low Pass &
\textbf{1.000} & \textbf{1.000} & \textbf{1.000} &
\textbf{1.000} & 0.984 & 0.784 &
\textbf{1.000} & 0.945 & 0.559 &
-- & \textbf{1.000} & \textbf{1.000} &
\textbf{1.000} & 0.999 & 0.982 \\

& High Pass &
\textbf{1.000} & \textbf{1.000} & \textbf{1.000} &
\textbf{1.000} & 0.933 & 0.369 &
\textbf{1.000} & 0.999 & 0.982 &
-- & 0.901 & 0.847 &
0.996 & \textbf{1.000} & \textbf{1.000} \\

& Equalization &
\textbf{1.000} & \textbf{1.000} & \textbf{1.000} &
0.991 & 0.890 & 0.243 &
\textbf{1.000} & 0.845 & 0.243 &
-- & 0.995 & 0.982 &
\textbf{1.000} & 0.993 & 0.892 \\

& Polarity Inversion &
\textbf{1.000} & \textbf{1.000} & \textbf{1.000} &
0.870 & 0.171 & 0.000 &
0.500 & 0.213 & 0.000 &
-- & \textbf{1.000} & \textbf{1.000} &
\textbf{1.000} & \textbf{1.000} & \textbf{1.000} \\

& Echo &
\textbf{1.000} & \textbf{1.000} & \textbf{1.000} &
\textbf{1.000} & 0.969 & 0.667 &
0.996 & 0.983 & 0.793 &
-- & \textbf{1.000} & \textbf{1.000} &
\textbf{1.000} & 0.999 & 0.991 \\

& Resample &
\textbf{1.000} & \textbf{1.000} & \textbf{1.000} &
\textbf{1.000} & 0.985 & 0.820 &
\textbf{1.000} & 0.996 & 0.937 &
-- & \textbf{1.000} & \textbf{1.000} &
\textbf{1.000} & 0.990 & 0.874 \\

\midrule
\multirow{5}{*}{\textbf{Temporal}}
& Cut &
\textbf{1.000} & \textbf{1.000} & \textbf{1.000} &
0.870 & 0.551 & 0.009 &
\textbf{1.000} & 0.506 & 0.000 &
-- & 0.791 & 0.748 &
\textbf{1.000} & \textbf{1.000} & \textbf{1.000} \\

& Time Stretch &
\textbf{1.000} & 0.996 & \textbf{0.964} &
0.500 & 0.595 & 0.000 &
0.514 & 0.617 & 0.009 &
-- & 0.349 & 0.000 &
\textbf{1.000} & \textbf{0.997} & \textbf{0.964} \\

& Time Jittering &
0.991 & 0.974 & 0.928 &
\textbf{1.000} & 0.980 & 0.748 &
\textbf{1.000} & 0.999 & 0.982 &
-- & 0.995 & 0.964 &
\textbf{1.000} & \textbf{1.000} & \textbf{1.000} \\

& Phase Shift &
\textbf{1.000} & \textbf{1.000} & \textbf{1.000} &
0.898 & 0.599 & 0.018 &
\textbf{1.000} & 0.498 & 0.000 &
-- & 0.855 & 0.812 &
\textbf{1.000} & 0.999 & 0.982 \\

& Speed &
0.523 & 0.045 & 0.045 &
0.500 & 0.497 & 0.000 &
\textbf{1.000} & 0.502 & 0.000 &
-- & 0.245 & 0.000 &
\textbf{1.000} & \textbf{0.978} & \textbf{0.685} \\

\midrule
\multirow{2}{*}{\textbf{Traditional Codec}}
& MP3 &
0.991 & 0.974 & 0.910 &
0.924 & 0.661 & 0.000 &
0.996 & \textbf{0.998} & \textbf{0.964} &
-- & 0.785 & 0.775 &
\textbf{1.000} & 0.994 & 0.910 \\

& AAC &
\textbf{1.000} & \textbf{1.000} & \textbf{1.000} &
\textbf{1.000} & 0.990 & 0.874 &
\textbf{1.000} & 0.996 & 0.955 &
-- & 0.995 & 0.973 &
\textbf{1.000} & 0.999 & 0.982 \\

\midrule
\multirow{3}{*}{\textbf{Neural Codec}}
& Encodec~\cite{defossez2022high} &
0.505 & 0.002 & 0.000 &
\textbf{1.000} & 0.900 & 0.180 &
0.982 & 0.530 & 0.000 &
-- & 0.261 & 0.000 &
\textbf{1.000} & \textbf{0.989} & \textbf{0.847} \\

& DAC~\cite{kumar2023high} &
0.505 & 0.004 & 0.000 &
0.996 & \textbf{0.948} & \textbf{0.523} &
0.572 & 0.615 & 0.000 &
-- & 0.471 & 0.009 &
\textbf{1.000} & 0.933 & 0.342 \\

& SNAC~\cite{siuzdak2024snac} &
0.500 & 0.000 & 0.000 &
0.500 & 0.525 & 0.000 &
0.505 & 0.471 & 0.000 &
-- & 0.276 & 0.000 &
\textbf{1.000} & \textbf{0.926} & \textbf{0.261} \\

\midrule
\multirow{2}{*}{\textbf{Music Specific}}
& Cover Song &
0.995 & 0.493 & 0.000 &
0.933 & 0.501 & 0.000 &
\textbf{0.999} & 0.989 & 0.838 &
-- & 0.992 & \textbf{0.967} &
0.998 & \textbf{0.994} & 0.938 \\

& Cover Song + Cut &
0.843 & 0.340 & 0.000 &
0.588 & 0.497 & 0.000 &
\textbf{0.999} & 0.504 & 0.005 &
-- & 0.427 & 0.127 &
0.993 & \textbf{0.980} & \textbf{0.813} \\

\bottomrule
\end{tabular}%
}
\end{table*}
\noindent\textbf{Implementation Details.}
We optimize all trainable parameters using AdamW with a learning rate of $1 \times 10^{-4}$, following a linear warmup of 500 steps and a linear annealing schedule decaying to $1 \times 10^{-5}$. We train for 60K steps with an effective batch size of 8, applying attack augmentation starting from the 8K step. All experiments are conducted in bf16 mixed precision on 8 NVIDIA L40S GPUs, totaling approximately 576 GPU hours. Detailed training settings are provided in Table~\ref{tab:training_details}.

\noindent\textbf{Attack Evaluation.}
We evaluate watermark robustness under 20 attacks, consisting of 18 general audio attacks and two music-specific cover-song attacks. The general audio attacks include common signal perturbations, temporal transformations, traditional codecs, and neural codecs, while the two music-specific attacks correspond to cover-song and cover-song with cut.
For the 18 general audio attacks, we use a 2K-sample evaluation set and uniformly sample one attack for each watermarked example, assigning approximately 111 examples to each attack on average. 

For the two cover-song attacks, we evaluate the same 2K samples using the pipeline shown in Fig. 2, where the vocal stem is converted to a different singer identity and optionally cropped. Details of the source-separation and voice-conversion configuration are provided in Appendix~\ref{app:attack_details}.

\begin{table*}[!t]
\caption{Objective evaluation of generation quality. CE, CU, PC, and PQ denote content enjoyment, content usefulness, production complexity, and production quality, respectively.}
\label{tab:objective_quality}
\centering
\scriptsize
\small
\setlength{\tabcolsep}{2.5pt}
\renewcommand{\arraystretch}{1.08}
\begin{tabular*}{\textwidth}{@{\extracolsep{\fill}}lccccccc@{}}
\toprule
\multirow{2}{*}{\textbf{Model}} &
\multicolumn{1}{c}{\textbf{Distribution}} &
\multicolumn{2}{c}{\textbf{Alignment}} &
\multicolumn{4}{c}{\textbf{Aesthetics}} \\
\cmidrule(lr){2-2}
\cmidrule(lr){3-4}
\cmidrule(l){5-8}
&
\textbf{FAD $\downarrow$} &
\textbf{CLAP $\uparrow$} &
\textbf{PER $\downarrow$} &
\textbf{CE $\uparrow$} &
\textbf{CU $\uparrow$} &
\textbf{PC $\uparrow$} &
\textbf{PQ $\uparrow$} \\
\midrule
\multicolumn{8}{c}{\textbf{Music Generation Baselines}} \\
\midrule
ACE-Step~\cite{gong2025ace} &
0.5582 & \textbf{0.2843} & 0.3395 & 6.6712 & 7.2959 & 6.2314 & 7.4071 \\
\midrule
\multicolumn{8}{c}{\textbf{Post-hoc Watermarking}} \\
\midrule
AudioSeal-M~\cite{sanroman2024proactive} &
\textbf{0.5505} & 0.2786 & 0.3537 & 6.6570 & 7.2737 & 6.2314 & 7.3771 \\
SilentCipher~\cite{singh2024silentcipher} &
0.6411 & 0.2675 & 0.3575 & 6.4821 & 7.1382 & 6.0507 & 7.3969 \\
\midrule
\multicolumn{8}{c}{\textbf{Generative Watermarking}} \\
\midrule
\textbf{MusicMark} &
0.5633 & \textbf{0.2843} & \textbf{0.3002} & \textbf{6.7432} & \textbf{7.3340} & \textbf{6.3393} & \textbf{7.4005} \\
\bottomrule
\end{tabular*}

\end{table*}
\noindent\textbf{Evaluation Metrics.} 
We evaluate MusicMark in terms of watermark robustness and generation quality. For watermark robustness, we report detection accuracy (Acc), computed on a balanced set of watermarked and unwatermarked samples to measure whether the detector correctly identifies watermark presence or absence, bit accuracy (Bit), which measures bit-wise message extraction accuracy, and absolute message accuracy (Abs), which requires all message bits to be correctly extracted. For generation quality, we report Fr\'echet Audio Distance (FAD)~\cite{kilgour2018fr} for music quality, CLAP score~\cite{elizalde2023clap} and Phoneme Error Rate (PER) for music-text condition alignment, and Audiobox-Aesthetic~\cite{tjandra2025meta} for aesthetic quality as objective metrics. We also conduct human evaluation using Mean Opinion Score (MOS) ratings on overall quality (OVL), text relevance (REL), vocal quality (VQ), and vocal-instrumental harmony (HAM). Details of the objective quality evaluation and human evaluation are provided in Appendix~\ref{app:eval_metric}.

\subsection{Watermark Robustness}
\noindent\textbf{Comparison with Post-hoc Watermarking.} 
Table~\ref{tab:attack_wm} shows the watermark robustness results. Under no-attack conditions, all models except AudioSeal~\cite{sanroman2024proactive} achieve near-perfect bit and absolute accuracy, demonstrating that watermarks can be reliably detected and extracted in the absence of attacks. However, under attack conditions, post-hoc baselines suffer a significant performance drop across all metrics, whereas MusicMark maintains strong robustness. This contrast is most evident in the comparison with AudioSeal-M, which is trained with the same data and attack augmentations as MusicMark. While both models perform well under no-attack conditions, AudioSeal-M drops sharply under attacks with an absolute accuracy of 0.400, compared to MusicMark's 0.869. This demonstrates that the robustness gap stems not from training conditions but from the watermarking approach itself. Since post-hoc methods insert watermarks after music is generated, the watermark remains fragile to diverse attacks, whereas MusicMark embeds watermarks into the semantic latent space during generation, achieving substantially stronger robustness.

\noindent\textbf{Standard Audio Attacks.}
In Table~\ref{tab:attack_wm}, standard audio attacks consist of the Common, Temporal, and Traditional Codec categories. MusicMark maintains consistently strong robustness across all attack types. Post-hoc baselines, in contrast, can achieve strong performance on specific attacks but lack consistent robustness across all attack types. For example, SilentCipher~\cite{singh2024silentcipher} achieves near-perfect absolute accuracy on Reverb and Equalization but completely fails under Time Stretch and Speed. AudioSeal-M similarly collapses under Cut, Phase Shift, and Speed. This inconsistency stems from the nature of post-hoc watermarking, which introduces imperceptible perturbations into already generated audio, making the watermark inherently fragile to diverse audio transformations. In contrast, since MusicMark embeds watermarks into the semantic latent space during generation, the watermark resides within the semantic representation of the musical content itself, achieving consistently strong robustness across all general audio attacks.

\begin{figure*}[t!]
    \centering
    \includegraphics[width=\textwidth]{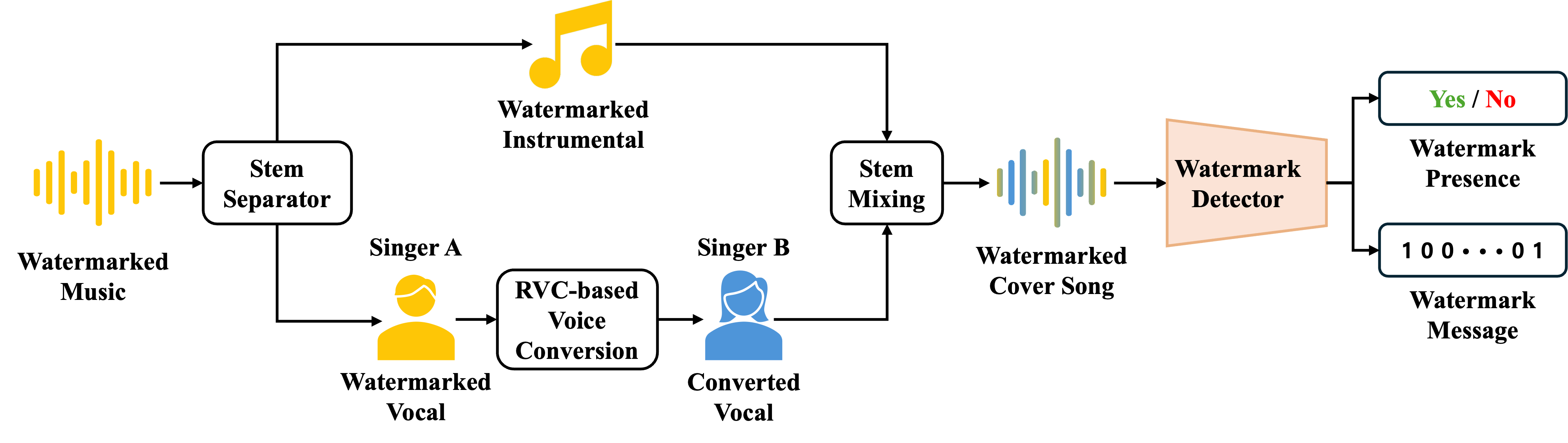}
    \caption{Overview of the cover-song attack pipeline. A watermarked music sample is separated into instrumental and vocal stems, and the vocal stem is converted to a different singer identity using an RVC-based voice conversion model. The converted vocal is then remixed with the watermarked instrumental to generate a watermarked cover song, which is passed to the watermark detector for presence detection and message extraction.}
    \label{fig2}
\end{figure*}

\noindent\textbf{Neural Codec Attacks.}
As shown in Table~\ref{tab:attack_wm}, neural codec re-synthesis is the most severe attack for post-hoc watermarking methods. Neural codecs reconstruct audio from compact learned representations that preserve perceptually salient content while suppressing fine-grained residual details. Since post-hoc watermarks are introduced as imperceptible perturbations rather than part of the semantic content, they are stripped away during this process. 
Across neural codec attacks, most post-hoc baselines suffer severe degradation, particularly in absolute message accuracy. WavMark, AudioSeal-M, and SilentCipher collapse to near-zero absolute accuracy across the three neural codecs, while AudioSeal remains partially robust under DAC but still degrades under EnCodec and SNAC. In contrast, MusicMark maintains perfect detection accuracy and substantially higher bit accuracy across all neural codec attacks.

\noindent\textbf{Cover-Song Attacks.} 
We further analyze robustness against music-specific cover-song attacks, whose evaluation pipeline is described in Fig.~\ref{fig2}. As shown in Table~\ref{tab:attack_wm}, under the Cover Song attack alone, MusicMark achieves 0.998 detection accuracy, 0.994 bit accuracy, and 0.938 absolute accuracy. The 44.1~kHz music-oriented methods remain largely robust, whereas the 16~kHz baselines are vulnerable in message extraction, with WavMark and AudioSeal dropping to 0.000 absolute accuracy.

The Cover Song + Cut attack is substantially more challenging. In recent online music platforms, songs are increasingly shared as short clips, and modified versions of existing music are also widely circulated. Such modifications are commonly created using readily available music-editing tools that alter the vocal component while preserving much of the original accompaniment. To reflect this realistic redistribution scenario, we evaluate watermark robustness under a combined attack that applies both cover-song style vocal modification and temporal cutting. Under this setting, the absolute accuracy of AudioSeal-M and SilentCipher drops to 0.005 and 0.127, respectively. In particular, SilentCipher achieves 0.748 absolute accuracy under Cut alone, but degrades to 0.127 when cut is combined with the cover-song transformation, showing that music-specific manipulation further exposes the vulnerability of post-hoc watermarking methods. In contrast, MusicMark maintains 0.813 absolute accuracy, with 0.993 detection accuracy and 0.980 bit accuracy. This demonstrates that embedding the watermark into the semantic latent space during generation provides stronger robustness against realistic music-specific manipulations.

\begin{figure*}[t!]
    \centering
    \includegraphics[width=\textwidth]{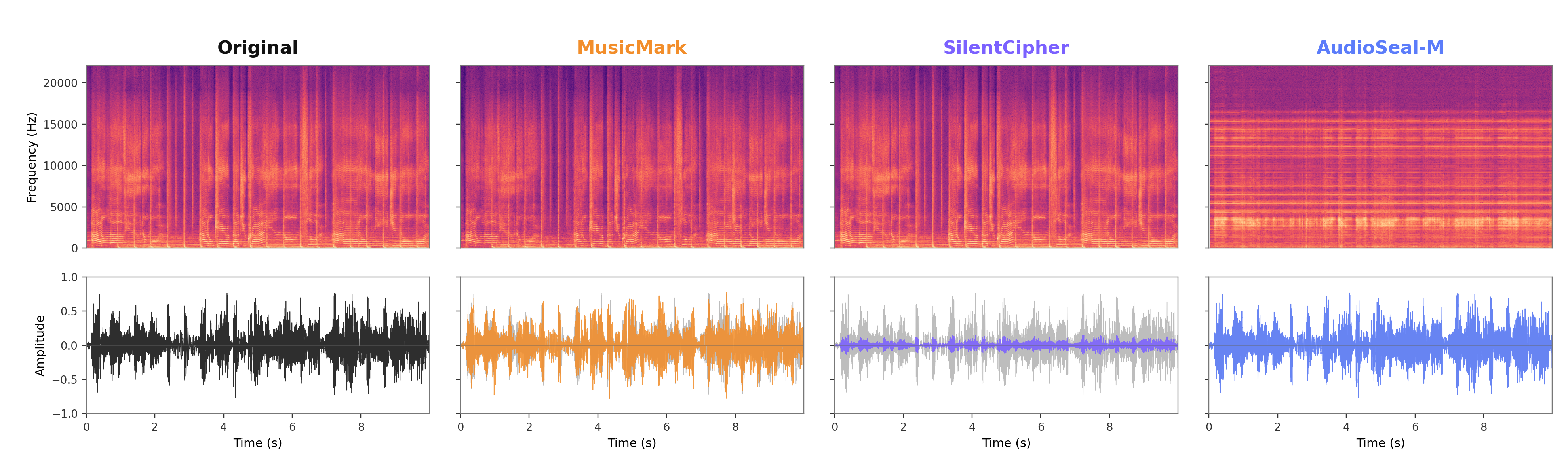}
    \caption{Qualitative comparison of music watermarking methods. 
The Original column shows an unwatermarked audio excerpt, and the remaining columns show the differences between each watermarked output and the original audio. The top row shows mel-spectrograms, and the bottom row shows 
waveforms, where the original waveform is shown in gray and each watermarked 
waveform is shown in color.}
    \label{fig3}
\end{figure*}

\subsection{Generation Quality}
\noindent\textbf{Objective Evaluation Results.}
As shown in Table~\ref{tab:objective_quality}, MusicMark preserves the quality of ACE-Step~\cite{gong2025ace} while adding watermarking capability. Compared with ACE-Step, MusicMark maintains comparable FAD and CLAP scores, reduces PER from 0.3395 to 0.3002, and achieves similar or slightly better aesthetic scores, showing that watermark injection does not degrade generation quality.

Compared with post-hoc watermarking baselines, MusicMark achieves better alignment and aesthetic scores while maintaining competitive FAD. This improvement is especially clear compared with AudioSeal-M, which is trained with the same data as MusicMark. Since post-hoc methods add watermark signals to already generated music, they can introduce quality degradation. In contrast, MusicMark embeds the watermark within semantic latent generation, preserving music quality more effectively.

\noindent\textbf{Qualitative Spectrogram Analysis.}
To qualitatively examine how different watermarking methods affect the acoustic structure of music, we visualize the watermark signatures in Fig.~\ref{fig3}. The Original column shows the mel-spectrogram and waveform of the unwatermarked music, while the other columns show the difference between each watermarked output and the original.

SilentCipher largely preserves the mel-spectrogram structure, but its waveform difference shows strong amplitude attenuation, indicating a noticeable loudness change. AudioSeal-M exhibits the opposite behavior: its waveform difference is relatively small, but the spectrogram difference reveals horizontal artifacts at fixed frequency bands, making the watermark signature visible in the frequency domain.

In contrast, MusicMark preserves the main time-frequency patterns of the original music without introducing fixed-band spectral artifacts as in AudioSeal-M or severe waveform attenuation as in SilentCipher. For qualitative inspection beyond the static spectrogram and waveform visualizations, we provide the corresponding music examples and additional samples on the demo page in Appendix~\ref{app:demo}.

\begin{table}[!t]
\caption{Human evaluation of generation quality}
\label{tab:human_eval}
\centering
\scriptsize
\setlength{\tabcolsep}{6pt}
\renewcommand{\arraystretch}{1.12}
\resizebox{\columnwidth}{!}{%
\begin{tabular}{@{}lcccc@{}}
\toprule
\multirow{2}{*}{\textbf{Model}} &
\multicolumn{4}{c}{\textbf{MOS}} \\
\cmidrule(l){2-5}
&
\textbf{OVL $\uparrow$} &
\textbf{REL $\uparrow$} &
\textbf{VQ $\uparrow$} &
\textbf{HAM $\uparrow$} \\
\midrule
ACE-Step & \textbf{3.64 $\pm$ 0.44} & 3.85 $\pm$ 0.36 & \textbf{3.86 $\pm$ 0.48} & \textbf{3.69 $\pm$ 0.37} \\
AudioSeal-M & 3.44 $\pm$ 0.72 & 3.60 $\pm$ 0.79 & 3.68 $\pm$ 0.81 & 3.53 $\pm$ 0.66 \\
SilentCipher & 2.74 $\pm$ 0.92 & 3.16 $\pm$ 0.92 & 3.20 $\pm$ 1.06 & 2.76 $\pm$ 1.01 \\
\textbf{MusicMark} & 3.62 $\pm$ 0.43 & \textbf{3.85 $\pm$ 0.46} & 3.79 $\pm$ 0.45 & 3.62 $\pm$ 0.40 \\
\bottomrule
\end{tabular}%
}
\end{table}
\noindent\textbf{Human Evaluation.}
We conduct human evaluation using MOS ratings to assess perceptual music quality. As shown in Table~\ref{tab:human_eval}, MusicMark achieves scores comparable to the base model ACE-Step, with identical text relevance and only marginal differences in the other metrics. In contrast, post-hoc watermarking baselines show consistent degradation across all metrics, with SilentCipher~\cite{singh2024silentcipher} exhibiting the largest drop, especially in overall quality and vocal-instrument harmony. These results confirm that the quality preservation observed in objective metrics is also reflected in human perception.

\FloatBarrier
\subsection{Ablation Study}
\label{sec:ablation}

To validate the key design choices of MusicMark, we conduct ablation studies from two perspectives. First, within our framework, we analyze four factors that affect the robustness-quality trade-off of watermark embedding during generation: the insertion stage, injection mechanism, adapter layer position, and latent consistency loss. Second, we instantiate MusicMark on a different music generation model to verify that the proposed framework is not tied to a specific architecture.

\begin{table*}[!t]
\caption{Ablation Study of MusicMark: Watermark Robustness and Generation Quality. Standard Attack includes common, temporal, and traditional codec attacks.}
\label{tab:ablation_all}
\centering
\tiny
\setlength{\tabcolsep}{2.6pt}
\renewcommand{\arraystretch}{1.12}
\resizebox{\textwidth}{!}{%
\begin{tabular}{@{}l *{15}{c}@{}}
\toprule
\multicolumn{16}{c}{\textbf{Watermark Robustness}} \\
\midrule
\multicolumn{1}{c}{\multirow{2}{*}{\textbf{Variant}}} &
\multicolumn{3}{c}{\textbf{No Attack}} &
\multicolumn{3}{c}{\textbf{Standard Attack}} &
\multicolumn{3}{c}{\textbf{Neural Codec}} &
\multicolumn{3}{c}{\textbf{Cover Song}} &
\multicolumn{3}{c}{\textbf{Cover Song + Cut}} \\
\cmidrule(lr){2-4} \cmidrule(lr){5-7} \cmidrule(lr){8-10} \cmidrule(lr){11-13} \cmidrule(l){14-16}
&
\textbf{Acc $\uparrow$} & \textbf{Bit $\uparrow$} & \textbf{Abs $\uparrow$} &
\textbf{Acc $\uparrow$} & \textbf{Bit $\uparrow$} & \textbf{Abs $\uparrow$} &
\textbf{Acc $\uparrow$} & \textbf{Bit $\uparrow$} & \textbf{Abs $\uparrow$} &
\textbf{Acc $\uparrow$} & \textbf{Bit $\uparrow$} & \textbf{Abs $\uparrow$} &
\textbf{Acc $\uparrow$} & \textbf{Bit $\uparrow$} & \textbf{Abs $\uparrow$} \\
\midrule
\multicolumn{1}{c}{\textbf{MusicMark}} &
\textbf{1.000} & 0.999 & 0.993 &
0.999 & \textbf{0.988} & 0.859 &
\textbf{1.000} & \textbf{0.950} & 0.486 &
0.998 & \textbf{0.994} & \textbf{0.938} &
0.993 & \textbf{0.980} & \textbf{0.813} \\
\midrule
\multicolumn{16}{c}{\textbf{Watermark Stage}} \\
\quad (a) Vocoder Stage &
\textbf{1.000} & \textbf{1.000} & \textbf{1.000} &
0.892 & 0.947 & 0.795 &
0.351 & 0.726 & 0.102 &
0.955 & 0.984 & 0.884 &
0.904 & 0.944 & 0.712 \\
\midrule
\multicolumn{16}{c}{\textbf{Injection Method}} \\
\quad (b) Shared Cross-attn &
\textbf{1.000} & 0.988 & 0.818 &
0.997 & 0.959 & 0.631 &
0.890 & 0.863 & 0.123 &
\textbf{0.999} & 0.978 & 0.657 &
\textbf{0.997} & 0.969 & 0.568 \\
\midrule
\multicolumn{16}{c}{\textbf{Layer Position}} \\
\quad (c) Middle-6 &
\textbf{1.000} & 0.999 & 0.993 &
\textbf{1.000} & 0.979 & 0.816 &
\textbf{1.000} & 0.907 & 0.273 &
0.970 & 0.989 & 0.915 &
0.930 & 0.962 & 0.727 \\
\quad (d) Middle-12 &
\textbf{1.000} & 0.999 & 0.983 &
0.995 & 0.965 & 0.756 &
0.844 & 0.866 & 0.135 &
0.983 & 0.981 & 0.818 &
0.951 & 0.947 & 0.621 \\
\midrule
\multicolumn{16}{c}{\textbf{Loss Design}} \\
\quad (e) w/o Latent Loss &
\textbf{1.000} & 0.999 & 0.994 &
\textbf{1.000} & 0.987 & \textbf{0.874} &
0.993 & 0.945 & \textbf{0.505} &
0.980 & 0.992 & 0.936 &
0.949 & 0.971 & 0.757 \\

\midrule
\midrule
\multicolumn{16}{c}{\textbf{Generation Quality}} \\
\midrule
\multicolumn{1}{c}{\multirow{2}{*}{\textbf{Variant}}} &
\multicolumn{3}{c}{\textbf{Distribution}} &
\multicolumn{4}{c}{\textbf{Alignment}} &
\multicolumn{8}{c}{\textbf{Aesthetics}} \\
\cmidrule(lr){2-4} \cmidrule(lr){5-8} \cmidrule(l){9-16}
&
\multicolumn{3}{c}{\textbf{FAD $\downarrow$}} &
\multicolumn{2}{c}{\textbf{CLAP $\uparrow$}} &
\multicolumn{2}{c}{\textbf{PER $\downarrow$}} &
\multicolumn{2}{c}{\textbf{CE $\uparrow$}} &
\multicolumn{2}{c}{\textbf{CU $\uparrow$}} &
\multicolumn{2}{c}{\textbf{PC $\uparrow$}} &
\multicolumn{2}{c}{\textbf{PQ $\uparrow$}} \\
\midrule
\multicolumn{1}{c}{\textbf{MusicMark}} &
\multicolumn{3}{c}{0.5633} &
\multicolumn{2}{c}{\textbf{0.2843}} &
\multicolumn{2}{c}{0.3002} &
\multicolumn{2}{c}{\textbf{6.7432}} &
\multicolumn{2}{c}{\textbf{7.3340}} &
\multicolumn{2}{c}{6.3393} &
\multicolumn{2}{c}{\textbf{7.4005}} \\
\midrule
\multicolumn{16}{c}{\textbf{Watermark Stage}} \\
\quad (a) Vocoder Stage &
\multicolumn{3}{c}{\textbf{0.5485}} &
\multicolumn{2}{c}{0.2272} &
\multicolumn{2}{c}{0.4073} &
\multicolumn{2}{c}{6.0224} &
\multicolumn{2}{c}{6.6416} &
\multicolumn{2}{c}{6.0768} &
\multicolumn{2}{c}{6.5709} \\
\midrule
\multicolumn{16}{c}{\textbf{Injection Method}} \\
\quad (b) Shared Cross-attn &
\multicolumn{3}{c}{0.5716} &
\multicolumn{2}{c}{0.2603} &
\multicolumn{2}{c}{0.8215} &
\multicolumn{2}{c}{6.4367} &
\multicolumn{2}{c}{7.0514} &
\multicolumn{2}{c}{\textbf{6.5201}} &
\multicolumn{2}{c}{6.8451} \\
\midrule
\multicolumn{16}{c}{\textbf{Layer Position}} \\
\quad (c) Middle-6 &
\multicolumn{3}{c}{0.5528} &
\multicolumn{2}{c}{0.2650} &
\multicolumn{2}{c}{0.3102} &
\multicolumn{2}{c}{6.6634} &
\multicolumn{2}{c}{7.2842} &
\multicolumn{2}{c}{6.2243} &
\multicolumn{2}{c}{7.4002} \\
\quad (d) Middle-12 &
\multicolumn{3}{c}{0.5685} &
\multicolumn{2}{c}{0.2651} &
\multicolumn{2}{c}{\textbf{0.2962}} &
\multicolumn{2}{c}{6.6623} &
\multicolumn{2}{c}{7.2838} &
\multicolumn{2}{c}{6.1947} &
\multicolumn{2}{c}{7.3910} \\
\midrule
\multicolumn{16}{c}{\textbf{Loss Design}} \\
\quad (e) w/o Latent Loss &
\multicolumn{3}{c}{0.5636} &
\multicolumn{2}{c}{0.2505} &
\multicolumn{2}{c}{0.3346} &
\multicolumn{2}{c}{6.6003} &
\multicolumn{2}{c}{7.2317} &
\multicolumn{2}{c}{6.2736} &
\multicolumn{2}{c}{7.3009} \\
\bottomrule
\end{tabular}%
}
\end{table*}
\begin{table}[!t]
\caption{Watermark robustness and generation quality of MusicMark applied to the SA3 backbone. SA3 denotes Stable Audio 3.}
\label{tab:ablation_sa3}
\centering
\scriptsize
\setlength{\tabcolsep}{2.5pt}
\renewcommand{\arraystretch}{1.15}
\resizebox{\columnwidth}{!}{%
\begin{tabular}{@{}lcccccc@{}}
\toprule
\multicolumn{7}{c}{\textbf{Watermark Robustness}} \\
\midrule
\textbf{Setting} &
\multicolumn{2}{c}{\textbf{Acc $\uparrow$}} &
\multicolumn{2}{c}{\textbf{Bit $\uparrow$}} &
\multicolumn{2}{c}{\textbf{Abs $\uparrow$}} \\
\midrule
SA3 + AudioSeal-M &
\multicolumn{2}{c}{0.911} &
\multicolumn{2}{c}{0.774} &
\multicolumn{2}{c}{0.444} \\

SA3 + SilentCipher &
\multicolumn{2}{c}{--} &
\multicolumn{2}{c}{0.786} &
\multicolumn{2}{c}{0.675} \\

\textbf{SA3 + MusicMark} &
\multicolumn{2}{c}{\textbf{0.998}} &
\multicolumn{2}{c}{\textbf{0.920}} &
\multicolumn{2}{c}{\textbf{0.720}} \\

\midrule
\multicolumn{7}{c}{\textbf{Generation Quality}} \\
\midrule
\textbf{Model} &
\textbf{FAD $\downarrow$} &
\textbf{CLAP $\uparrow$} &
\textbf{CE $\uparrow$} &
\textbf{CU $\uparrow$} &
\textbf{PC $\uparrow$} &
\textbf{PQ $\uparrow$} \\
\midrule
SA3~\cite{evans2026stable} &
\textbf{0.2183} & \textbf{0.3202} & 6.7275 & 7.4778 & 5.6736 & 7.5166 \\

SA3 + AudioSeal-M &
0.2193 & 0.3128 & 6.6968 & 7.4385 & 5.6739 & 7.4593 \\

SA3 + SilentCipher &
0.3250 & 0.3129 & 6.7346 & \textbf{7.4934} & 5.6091 & \textbf{7.5856} \\

\textbf{SA3 + MusicMark} &
0.2342 & 0.3126 & \textbf{6.7347} & 7.3691 & \textbf{5.7958} & 7.3374 \\

\bottomrule
\end{tabular}%
}
\end{table}
\noindent\textbf{Effect of Watermark Stage.}
To examine the importance of watermarking during semantic latent generation, we compare MusicMark with a vocoder-stage baseline that injects the watermark after the semantic representation has already been formed. In this baseline, following~\cite{liu2026vocbulwark}, the watermark representation is repeated along the temporal dimension and inserted into the first upsampling layer of the vocoder. As shown in Table~\ref{tab:ablation_all}(a), the vocoder-stage baseline performs well without attacks, but its robustness degrades under attacked settings. Specifically, this degradation is observed under standard attacks and becomes severe under neural codec attacks, where its absolute accuracy drops to 0.102. It also remains weaker than MusicMark under Cover Song + Cut. For generation quality, the vocoder-stage baseline obtains competitive FAD but substantially degrades CLAP, PER, and aesthetic scores, indicating weaker text-music alignment and perceptual quality. These results show that inserting the watermark after semantic generation is insufficient for robust music watermarking that preserves text-music alignment, while semantic latent watermarking provides stronger robustness and better quality preservation.

\noindent\textbf{Effect of Injection Method.}
We study how to effectively inject watermark information into semantic latent generation through cross-attention. As a baseline, \emph{Shared Cross-Attention} directly concatenates the watermark representation with the encoded lyric and style condition embeddings and feeds the merged sequence into the original cross-attention module, whereas MusicMark uses a decoupled cross-attention module. As shown in Table~\ref{tab:ablation_all}(b), direct concatenation already weakens message extraction under the no-attack setting and becomes substantially more vulnerable under neural codec and cover-song attacks. Its absolute accuracy drops to 0.123 under neural codec attacks and 0.568 under Cover Song + Cut, compared with 0.486 and 0.813 for MusicMark. The generation quality results further show that this misalignment affects lyric intelligibility, with PER increasing from 0.3002 to 0.8215. These results show that decoupled cross-attention provides a more stable way to inject watermark information into semantic generation while preserving both robustness and generation quality.

\noindent\textbf{Effect of Layer Position.}
We study which layers are most effective for inserting watermark adapters in the semantic latent generation model. In diffusion-based music generation models, earlier layers are often associated with lower-level refinements, while later layers can more directly affect the final semantic representation~\cite{gong2025ace}. Therefore, the adapter position can affect both watermark robustness and generation quality. We compare MusicMark's last-six-layer insertion with two mid-layer variants, Middle-6 (layers 10--15) and Middle-12 (layers 7--18), in the 24-layer generation model. As shown in Table~\ref{tab:ablation_all}(c)--(d), Middle-6 achieves stronger robustness than Middle-12, while both mid-layer variants remain inferior to MusicMark. The lower robustness of Middle-12 suggests that distributing adapter insertion over a broader middle-layer range makes the watermark less consistently reflected in the final semantic representation. The quality results show a similar trend: both mid-layer variants have lower CLAP and aesthetic scores than MusicMark. These results indicate that inserting adapters into the last layers more effectively embeds watermark information while preserving music quality.

\noindent\textbf{Effect of Latent Loss.}
The latent consistency loss is introduced to keep the watermark-conditioned denoised latent close to the corresponding unwatermarked latent under the same noise and text conditions. We evaluate its role using the w/o Latent Loss variant in Table~\ref{tab:ablation_all}(e). Without this constraint, the model can use the latent space more flexibly, which slightly improves absolute accuracy under standard and neural codec attacks. However, this can also perturb content-related latent information, reducing generation quality and robustness to music-specific transformations. The variant without latent loss drops from 0.813 to 0.757 absolute accuracy under Cover Song + Cut, and its CLAP score decreases substantially from 0.2843 to 0.2505, indicating weaker text-music alignment. It also obtains lower aesthetic scores across all dimensions. These results show that the latent consistency loss provides a better robustness-quality trade-off by preserving content-related latent structure while maintaining watermark robustness.

\noindent\textbf{Generalization of Music Generation Backbone.}
We examine whether the proposed framework can generalize across different diffusion-based music generation backbones by applying it to Stable Audio 3 (SA3)~\cite{evans2026stable}, a recent text-to-music diffusion model. We keep the SA3 backbone frozen and train only the watermark adapter and detector, using the same message length, training objective, and attack-augmentation pool as in the main experiments. Unlike ACE-Step, SA3 conditions only on style tags without lyrics and can generate instrumental music without vocals. We therefore exclude cover-song attacks, which require a vocal stem for conversion, and report average robustness over the standard and neural codec attacks.
As shown in Table~\ref{tab:ablation_sa3}, the SA3 + MusicMark keeps generation quality close to the unwatermarked backbone while outperforming the post-hoc baselines in all robustness metrics under attacks. Its quality scores remain comparable to the original SA3 output, indicating that the proposed framework does not substantially degrade the backbone's generation quality. These results suggest that the proposed framework can be transferred to other diffusion-based music generation models with a different conditioning scheme, indicating that it is not tied to a specific backbone architecture.
\section{Conclusion}
We propose MusicMark, the first generative watermarking framework tailored for lyrics- and text-conditioned music generation. MusicMark embeds watermark messages into the semantic latent space during diffusion-based music generation through a watermark adapter with decoupled cross-attention. By keeping the base generation model frozen and training only the adapter and detector, MusicMark couples watermarks with musical semantics while preserving the original generation capability. In addition, a latent consistency loss is introduced to better balance watermark robustness and generation quality by aligning the watermarked latent with its corresponding unwatermarked latent. Experimental results show that MusicMark significantly improves the robustness of watermark detection and message extraction over post-hoc watermarking baselines under diverse audio transformations, including neural codec re-synthesis and music-specific cover-song attacks, while maintaining perceptual quality. These results demonstrate that semantic latent watermarking during generation provides a practical and effective approach to provenance verification in AI-generated music. Current limitations include the 16-bit message capacity and evaluation mainly in text-conditioned music generation settings. Future work could explore higher-capacity watermarking, broader input conditions such as vocal and melody conditioning, and stronger generalization to more diverse and unseen music manipulations.

\bibliographystyle{IEEEtran}
\bibliography{references}

\appendix
\subsection{Implementation Details}
\label{app:imple_detail}
\noindent\textbf{MusicMark.}
MusicMark is implemented on ACE-Step~\cite{gong2025ace}, which consists of a Deep Compression AutoEncoder (DCAE), a diffusion transformer, and a vocoder; the backbone is frozen throughout training.

The watermark adapter first embeds each 16-bit message into 256-dimensional vectors. The embeddings are passed through $f_{\mathrm{Proj}}$, composed of two Linear--ReLU blocks with hidden dimensions 512 and 1024 and a final linear layer, followed by RMSNorm to obtain $\mathbf{Z}$. Key/value projections are zero-initialized with a learnable scale of 0.1, and the adapter is inserted into the last six of 24 cross-attention layers.

The detector feature extractor $\phi_{\mathrm{det}}$ is initialized from the pretrained DCAE encoder and applied to the log-mel spectrogram of the stereo waveform. We reuse the DCAE input convolution and first three encoder blocks, and use a prediction head $h_{\mathrm{pred}}$ to reduce the features to 32 channels, collapse the frequency dimension, and project them to $2+N$ presence and message logits. During training, each music instance is paired with a random 16-bit message. Attack augmentation starts after the first 8K steps to allow the adapter and detector to first learn reliable watermark embedding and detection in a clean setting before being exposed to diverse distortions. Each batch uses one attack sampled uniformly from the augmentation pool. Training settings are reported in Table~\ref{tab:training_details}.

\noindent\textbf{AudioSeal-M.}
AudioSeal-M adapts AudioSeal~\cite{sanroman2024proactive} to 44.1 kHz stereo music, retaining the original encoder backbone with strides [8, 5, 4, 2] and only modifying the input/output channel dimension from mono to stereo.
The input/output convolutions are extended to two channels by copying pretrained mono weights to both channels, with remaining layers initialized from the 16\,kHz checkpoint. We additionally rescale the frequency-dependent time-frequency loudness loss to the 44.1\,kHz Nyquist band, using $48$ Mel bands instead of the original $16$. AudioSeal-M is trained on the same data and with the same attack-augmentation pool using a batch size of $16$ for $100$ epochs, with all remaining settings following AudioSeal.

\subsection{Datasets}
\label{app:dataset}

\noindent\textbf{Training Dataset.} We construct the training set from Muse~\cite{jiang2026muse}, a large-scale dataset of approximately 116K synthetic songs with paired lyrics and style prompts. Since MusicMark conditions on both lyrics and style tags, we retain English-language samples that include both fields. To improve consistency between the textual condition and the music, we further filter the data using the style--music similarity score provided in the metadata and retain samples with a score of at least 0.5. From the filtered tracks, we uniformly sample 30--60-second clips as training instances. The resulting training set contains 50,000 English music--text pairs, totaling approximately 625 hours of music.

\noindent\textbf{Evaluation Dataset.}
For evaluation, we retain English entries from a public Suno metadata dataset with structural lyric tags, style prompts, and durations of 60–240 seconds, yielding 2,000 conditions for robustness evaluation. For generation-quality evaluation, we sample 300 conditions from this set while preserving duration and prompt diversity.

\begin{table*}[!t]
\caption{Training settings of MusicMark}
\label{tab:training_details}
\centering
\footnotesize
\setlength{\tabcolsep}{8pt}
\renewcommand{\arraystretch}{1.08}
\begin{tabular}{@{}p{0.28\textwidth}p{0.58\textwidth}@{}}
\toprule
\textbf{Parameter} & \textbf{Specification} \\
\midrule
Base Model & ACE-Step-v1-3.5B \\
Training Dataset & Muse dataset: 50,000 English-language clips; 625 hours \\
Training Steps & 60,000 \\
Local Batch Size & 1 \\
GPU & 8 $\times$ L40S 48GB \\
Mixed Precision & Bfloat16 \\
Optimizer & AdamW \\
Optimizer Hyperparameters & $\beta_1=0.8,\ \beta_2=0.9,\ \epsilon=10^{-8}$ \\
Learning Rate & $10^{-4}$ \\
Minimum Learning Rate & $10^{-5}$ \\
Learning Rate Scheduler & Linear warmup, linear annealing \\
Warm-up Steps & 500 \\
Weight Decay & 0.01 \\
Gradient Clipping & Global norm, 1.0 \\
Scale Factor & $\alpha=0.1$ \\
Adapter Position & Last 6 of 24 layers \\
Loss Weights & $\lambda_{\mathrm{flow}}{=}1.0,\ \lambda_{\mathrm{latent}}{=}0.5,\ \lambda_{\mathrm{det}}{=}0.5,\ \lambda_{\mathrm{msg}}{=}1.0,\ \lambda_{1}{=}0.1,\ \lambda_{\mathrm{msspec}}{=}0.1,\ \lambda_{\mathrm{msstft}}{=}0.05$ \\
\bottomrule
\end{tabular}
\end{table*}
\begin{table*}[t]
\centering
\small
\caption{Attack configurations used for training augmentation and evaluation. A dash denotes unused attacks}
\label{tab:attack_details}
\setlength{\tabcolsep}{4pt}
\renewcommand{\arraystretch}{1.15}
\begin{tabular}{l l l c c}
\toprule
\textbf{Category} & \textbf{Type} & \textbf{Parameter} & \textbf{Training Config} & \textbf{Evaluation Config} \\
\midrule
\multirow{8}{*}{Common}
& Gaussian Noise & SNR (dB) & $[35, 60]$ & $[20, 40]$ \\
& Reverb & SNR (dB) & -- & $[0, 6]$ \\
& Low-pass Filter & Cutoff (Hz) & $[4000, 16000]$ & $[3500, 6000]$ \\
& High-pass Filter & Cutoff (Hz) & $[50, 500]$ & $[250, 500]$ \\
& Equalization & Max gain (dB) & -- & $\pm[0.375, 0.75]$ \\
& Polarity Inversion & N/A & -- & N/A \\
& Echo & Volume / delay (s) & $[0.1, 0.4]$ / $[0.1, 0.3]$ & $[0.1, 0.4]$ / $[0.1, 0.3]$ \\
& Resample & Rate (Hz) & $\{16000, 24000, 48000\}$ & $\{16000, 24000, 48000\}$ \\
\midrule
\multirow{5}{*}{Temporal}
& Cut & Ratio & -- & $0.5$ \\
& Time Stretch & Rate & -- & $[0.75, 0.95] \cup [1.05, 1.25]$ \\
& Time Jittering & Scale & -- & $[0.20, 0.50]$ \\
& Phase Shift & Offset (s) & -- & $[-0.10, -0.05] \cup [0.05, 0.10]$ \\
& Speed & Factor & $[0.8, 1.2]$ & $[0.8, 0.99] \cup [1.01, 1.2]$ \\
\midrule
\multirow{2}{*}{Traditional Codec}
& MP3 & Bitrate (kbps) & $\{64, 128\}$ & $64$ \\
& AAC & Bitrate (kbps) & -- & $64$ \\
\midrule
\multirow{3}{*}{Neural Codec}
& EnCodec & Codebooks & $\{4, 8, 16, 32\}$ & $16$ \\
& DAC & Codebooks & -- & $\{7, 8\}$ \\
& SNAC & Codebooks & -- & 4 \\
\midrule
\multirow{2}{*}{Music Specific}
& Cover Song & VCTK Checkpoint & 10 Voices (7F/3M) & 10 Voices (7F/3M) \\
& Cover Song + Cut & Ratio & $0.5$ & $0.5$ \\
\bottomrule
\end{tabular}
\end{table*}
\subsection{Attack Configurations}
\label{app:attack_details}
Attacks are organized into five categories: Common, Temporal, Traditional Codec, Neural Codec, and Music-specific attacks. We describe each attack below and summarize the corresponding parameter ranges and configurations in Table~\ref{tab:attack_details}.

\noindent\textbf{Common Attacks.}
This category covers common signal-level distortions that audio routinely undergoes in practice.
\begin{itemize}
\item \textbf{Gaussian Noise:} Adds Gaussian noise to the waveform.
\item \textbf{Reverb:} Applies reverberation to emulate room acoustics~\cite{jeub2009binaural}.
\item \textbf{Low-pass Filter:} Removes frequencies above a cutoff.
\item \textbf{High-pass Filter:} Removes frequencies below a cutoff.
\item \textbf{Equalization:} Applies a frequency-dependent gain.
\item \textbf{Polarity Inversion:} Inverts the waveform polarity by multiplying it by $-1$.
\item \textbf{Echo:} Adds a delayed, attenuated copy of the signal.
\item \textbf{Resample:} Resamples the audio to an intermediate sampling rate and then back to the original rate.
\end{itemize}

\noindent\textbf{Temporal Attacks.}
This category alters the temporal structure of the audio by changing its timing, duration, or playback rate.
\begin{itemize}
\item \textbf{Cut:} Crops the audio to a contiguous segment of half its duration, starting from a randomly selected point.
\item \textbf{Time Stretch:} Rescales the time axis without changing pitch.
\item \textbf{Time Jittering:} Applies local perturbations to the time axis of the waveform.
\item \textbf{Phase Shift:} Shifts the waveform phase in time.
\item \textbf{Speed:} Changes the playback speed, altering both duration and pitch.
\end{itemize}

\noindent\textbf{Traditional Codec Attacks.}
This category applies conventional lossy codecs used for audio transmission and storage.
\begin{itemize}
\item \textbf{MP3:} Applies MP3 encode--decode compression.
\item \textbf{AAC:} Applies AAC encode--decode compression.
\end{itemize}

\noindent\textbf{Neural Codec Attacks.}
This category uses neural audio codecs to compress and reconstruct the waveform.
\begin{itemize}
\item \textbf{EnCodec:} Compresses and reconstructs the audio with the 24\,kHz EnCodec model~\cite{defossez2022high}.
\item \textbf{DAC:} Compresses and reconstructs the audio with the Descript Audio Codec~\cite{kumar2023high} at 44.1\,kHz.
\item \textbf{SNAC:} Compresses and reconstructs the audio with the multi-scale SNAC codec~\cite{siuzdak2024snac}.
\end{itemize}
\noindent\textbf{Music-specific Attacks.}
This category includes cover-song attacks that change the vocal identity while preserving the underlying song content.
\begin{itemize}
\item \textbf{Cover Song:} 
Separates the audio into vocal and instrumental stems using MDX-Net~\cite{kim2021kuielabmdxnet}, converts the vocal stem using an RVC-based voice conversion model\footnote{\url{https://github.com/RVC-Project/Retrieval-based-Voice-Conversion}} to one of 10 VCTK voice identities (7 female, 3 male) provided by a pretrained checkpoint\footnote{\url{https://huggingface.co/Nekochu/RVC-VCTK_Voice-sample}}, and remixes the converted vocal with the original instrumental stem.

\item \textbf{Cover Song + Cut:} Applies Cut after the cover-song transformation to simulate short-form redistribution.
\end{itemize}

\subsection{Evaluation Metrics}
\label{app:eval_metric}
\noindent\textbf{Objective Quality Evaluation.}
FAD is computed with the CLAP-LAION-Music embedding model using the \texttt{fadtk} package\footnote{\url{https://github.com/microsoft/fadtk}}, with the 4,830-track FMA-Pop subset as the reference distribution and each method's 300 generated outputs as the evaluated distribution. PER is measured by transcribing each generated sample with Whisper-large-v3~\cite{radford2023robust}, normalizing the transcription and ground-truth lyrics, converting both into phoneme sequences with \texttt{g2p\_en}, and computing reference-normalized edit distance. CLAP is measured as the cosine similarity between the style-prompt embedding and generated-audio embedding using \mbox{\texttt{stable-audio-metrics}}\footnote{\url{https://github.com/Stability-AI/stable-audio-metrics}} with the CLAP-LAION-Music model. Audiobox-Aesthetic~\cite{tjandra2025meta} provides content enjoyment (CE), content usefulness (CU), production complexity (PC), and production quality (PQ) scores.

\noindent\textbf{Human Evaluation.} Human evaluation is conducted on 20 lyric--prompt conditions randomly sampled from the generation-quality evaluation set. For each condition, ten annotators rate four anonymized and randomly ordered outputs--ACE-Step, AudioSeal-M, SilentCipher, and MusicMark--on a 5-point MOS scale for overall quality (OVL), text relevance (REL), vocal quality (VQ), and vocal-instrumental harmony (HAM). Final scores are averaged across annotators and conditions.

\subsection{Demo Page}
\label{app:demo}
Music samples comparing watermarked outputs (SilentCipher, AudioSeal-M, and MusicMark) against the unwatermarked reference, along with cover-song attacked outputs, are available at \url{https://anonymous.4open.science/w/MusicMark-58F6/}.

\end{document}